\documentclass[reprint, amsmath,amssymb,aps,prl]{revtex4-1}

\usepackage{graphicx}
\usepackage{dcolumn}
\usepackage{bm}

\begin{document}


\title{Quantum spin-liquid behavior in the spin-1/2 random-bond Heisenberg antiferromagnet on the kagome lattice}

\author{Hikaru Kawamura}
 \email{kawamura@ess.sci.osaka-u.ac.jp}
\author{Ken Watanabe}
\author{Tokuro Shimokawa}
\affiliation{
Department of Earth and Space Science, Graduate School of Science, Osaka University, Toyonaka 560-0043, Japan
}%

\date{\today}

\begin{abstract}
The effect of the quenched bond-randomness on the ordering of the $S=1/2$ antiferromagnetic Heisenberg model on the kagome lattice is investigated by means of an exact-diagonalization method. When the randomness exceeds a critical value, the ground state of the model exhibits a transition within the non-magnetic state into the randomness-relevant gapless spin-liquid state. Implications to the $S=1/2$ kagome-lattice antiferromagnet herbertsmithite is discussed.
\end{abstract}

\maketitle

  Since the proposal by P.W. Anderson of the resonating valence bond state \cite{Anderson}, the quantum spin-liquid (QSL) state possibly realized in certain $S=1/2$ frustrated magnets has attracted interest of researchers \cite{Balents}. After a long experimental quest, several candidate materials were recently reported in geometrically frustrated magnets, including both the triangular-lattice and the kagome-lattice antiferromagnets (AFs).

 Examples of the triangular-lattice AF might be $S=1/2$ organic salts such as $\kappa$-(ET)$_2$Cu$_2$(CN)$_3$ \cite{Shimizu,SYamashita,MYamashita,Jawad} and EtMe$_3$Sb[Pd(dmit)$_2$]$_2$ \cite{Itou,MYamashita2,SYamashita2,Jawad2}, which exhibit no magnetic ordering, neither regular nor random, down to a very low temperature. The QSL state of these compounds exhibits gapless behaviors characterized by, {\it e.g.\/}, the low-temperature specific heat \cite{SYamashita,SYamashita2} or the thermal conductivity \cite{MYamashita,MYamashita2} linear in the absolute temperature $T$. An example of the second category, the kagome-lattice AF, might be herbertsmithite CuZn$_3$(OH)$_6$Cl$_2$ \cite{Shores,Helton07,deVries08,Mendelse,Helton10,Freedman,Han}. No magnetic order is observed down to 20mK in spite of the exchange coupling of 180K, while gapless behaviors are observed in various physical quantities. 

 The origin of the QSL behavior observed in these triangular-lattice organic salts and the kagome-lattice herbertsmithite has attracted tremendous interest, and various theoretical proposals have been made. The simplest possible reference model might be the $S=1/2$ AF Heisenberg model with the nearest-neighbor (n.n.) bilinear coupling. In the triangular case, the ground state of such n.n. model is known to be the standard AF long-range order (LRO), the 120-degrees structure \cite{Bernu,Capriotti}. Hence, to explain the QSL behavior observed in these organic salts, some ingredient  not taken into account in the simplest Heisenberg model is required. In the kagome case, by contrast, the ground state of the n.n. model is believed to be some sort of QSL state without the magnetic LRO, although its nature is still under hot debate. Various scenarios, including the Z$_2$ spin liquid \cite{Sachdev,Lu,Yan,Depenbrock,Jiang}, the algebraic U(1) spin liquid \cite{Hastings,Ran,Nakano,Iqbal13}, the chiral spin liquid \cite{Messio} and the valence bond crystal (VBC) \cite{Marston,Singh08,Evenbly} {\it etc.\/} have been proposed. It is not entirely clear at the present stage, however, which of these, or any of these, applies to the experimentally observed QSL state of herbertsmithite.

 Most of the recent theories pre-assumes that the system is clean enough that the quenched randomness plays a negligible role. Meanwhile, it was recently suggested in ref.\cite{Watanabe} that the QSL of the triangular-lattice  organic salts might be the randomness-induced one, the random-singlet \cite{Dasgupta,Bhatt,Fisher} or the valence-bond glass (VBG) state \cite{Tarzia,Singh10}, in which the randomness is self-generated for the spin degrees of freedom at low temperatures via the random freezing of the electric-polarization degrees of freedom \cite{Jawad,Jawad2}. Such a randomness-induced QSL picture apparently explains many of the experimentally observed features of the triangular organic salts \cite{Watanabe}.

 In case of herbertsmithite, it has been realized from the earlier stage of research that certain amount of quenched randomness exists originated from the random substitution of magnetic Cu$^{2+}$ by non-magnetic Zn$^{2+}$ or vice versa. It was suggested initially that the substitution occurred on the kagome plane, providing the site-dilution for Cu$^{2+}$ on the kagome layer \cite{deVries08,Mendelse}. More recently, it was suggested that 15\% of Zn$^{2+}$ on the adjacent triangular layer were randomly substituted by Cu$^{2+}$, keeping the kagome layer intact \cite{Freedman}. This might give rise to the Cu$^{2+}$[kagome]-Cu$^{2+}$[triangular]-Cu$^{2+}$[kagome] exchange path, the kagome-triangular path being ferromagnetic.

 The effect of the site-dilution randomness was analyzed by Singh by the series-expansion method \cite{Singh10}. Starting with the VBC picture for the non-random kagome AF \cite{Singh08}, he argued that the random dilution induced the VBG state with gapless behaviors. In view of the work by Singh on the site-random kagome AF and of the recent experimental finding that the randomness in herbertsmithite might primarily of the bond-random-type, and also of the recent work of ref.\cite{Watanabe} on the bond-random triangular AF, we wish to investigate the effect of the bond-randomness on the ordering of the $S=1/2$ kagome Heisenberg AF. 

 We consider the AF bond-random $S=1/2$ Heisenberg model on the kagome lattice whose Hamiltonian is given by
\begin{equation}
 {\cal H}=\sum_{<ij>} J_{ij}\vec S_i\cdot \vec S_j - H \sum_i S_i^z ,
\end{equation}
where $\vec S_i=(S_i^x,S_i^y,S_i^z)$ is a spin-1/2 operator at the $i$-th site on the lattice, and $H$ is the magnetic-field intensity. For the exchange coupling $J_{ij}$, we assume for simplicity the random n.n. AF coupling obeying the bond-independent uniform distribution between [$(1-\Delta )J$, $(1+\Delta )J$] with the mean $J$. The parameter $\Delta$ represents the extent of the randomness: $\Delta=0$ corresponds to the regular system and $\Delta=1$ to the maximally random system.

 In herbertsmithite, the randomness for the kagome-layer Cu$^{2+}$ would come primarily from the interlayer exchange path, adding an effective ferromagnetic coupling to the original AF coupling on the kagome layer, somewhat reducing the magnitude of the AF coupling in the bond-random manner. Since an apical Cu$^{2+}$ on the triangular layer contributes to three Cu$^{2+}$ n.n. bonds on the kagome layer, 15\% Cu$^{2+}$ substitution on the triangular layer affects significant amount of the kagome-layer Cu$^{2+}$ bonds. Although the substituted Cu$^{2+}$ eventually couples different kagome layers giving rise to the three-dimensional (3D) coupling, it has been established experimentally that herbertsmithite behaves as a 2D system without the 3D order. Hence, we choose to consider the effect of the randomly substituted Cu$^{2+}$ somewhat indirectly, only within the 2D kagome model. Note that, in our modelling, the direct contribution from the interlayer Cu$^{2+}$ themselves is neglected.

 Thus, we adopt here the simplest form of the bond-randomness, with the aim of getting insight into new qualitative features possibly induced by the bond randomness.  By means of an exact diagonalization (ED) method,  both the zero- and finite-temperature properties of the model are computed for finite lattices. Another reason of our choice of the randomness is a computational one, {\it i.e.\/}, this form of randomness turns out to yield rather stable results even for very small sizes accessible by the ED method. The ED technique meets a difficulty for certain types of randomness, {\it e.g.\/}, the site-dilution randomness, where small system sizes hamper a systematic study of the size dependence keeping the randomness extent (hole concentration) fixed. 

 The total number of spins is $N=12-30$ for $T=0$, and $N=12,15$ and 18 for $T>0$, periodic boundary conditions being employed. Sample average is taken over 100 ($N=12,16,18,21,24$) 50 ($N=27$) and 12 ($N=30$) independent bond realizations in the $T=0$ calculation, while  100 ($N=12,15$) and 20 ($N=18$) in the $T>0$ calculation. In what follows, the energy, temperature and magnetic field are normalized in units of $J$.

\begin{figure}
 \includegraphics[width=8cm]{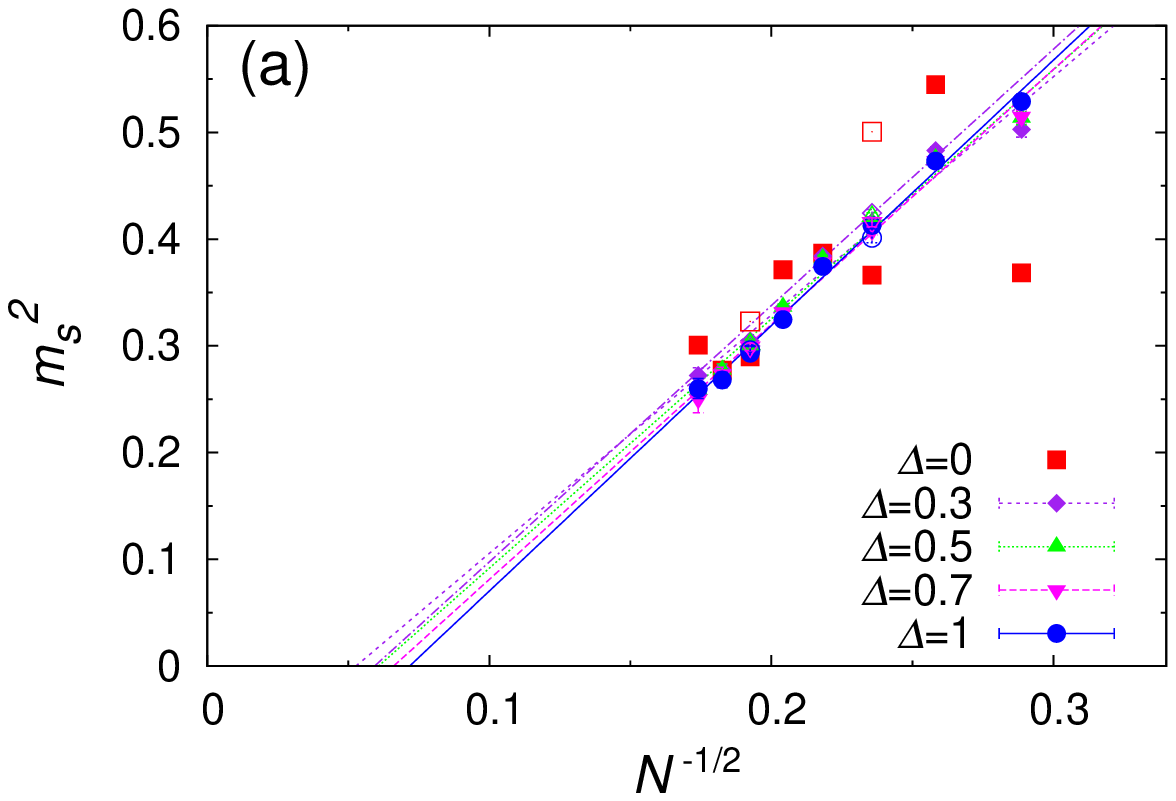}
 \includegraphics[width=8cm]{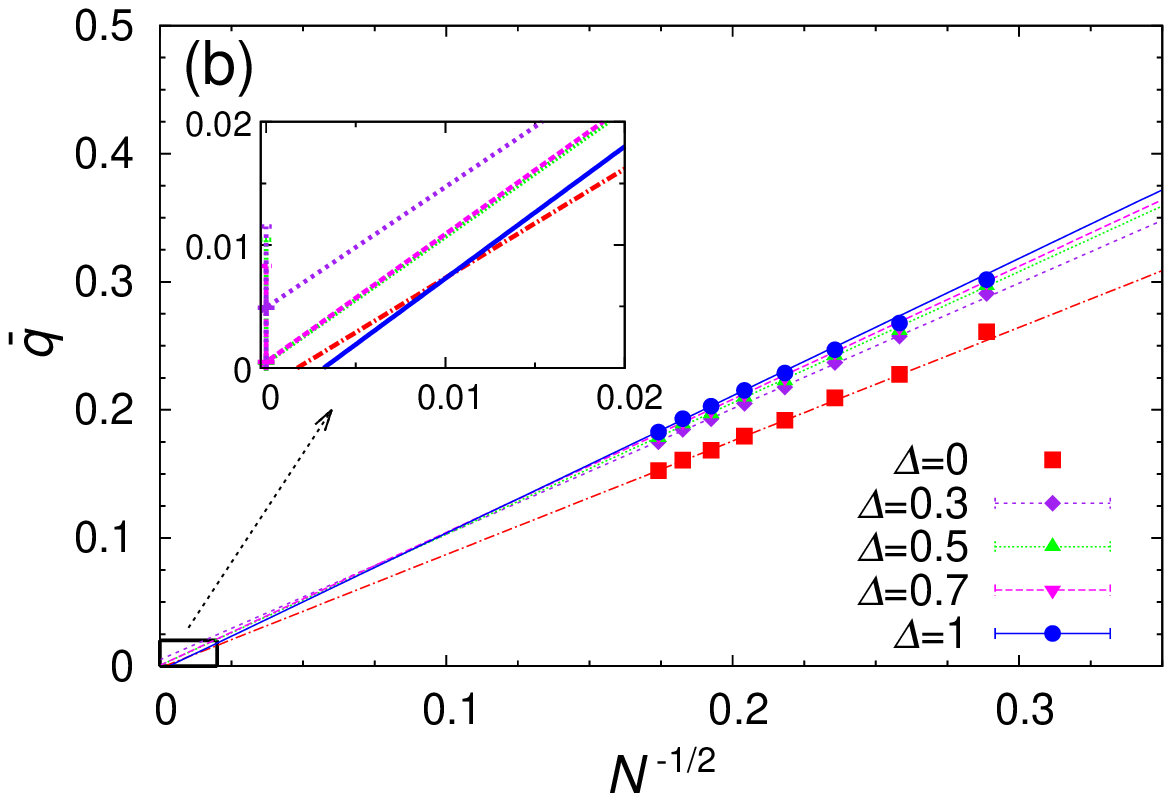}
 \caption{(color online). 
 (a) The squared sublattice magnetization $m_s^2$ associated with the three sublattices (solid symbols) and with the nine sublattices (open symbols, for $N=18,27$ only) for several randomness $\Delta$. The lines are linear fits of the data. (b) The spin freezing parameter $\bar q$ versus $1/\sqrt{N}$. The lines are the linear fits using the four largest-$N$ data. The inset is a magnified view of the large-$N$ region.
}
\end{figure}

 We first examine the existence or nonexistence of the magnetic LRO at $T=0$, by computing the squared sublattice magnetization $m_s^2$. We consider the three sublattices associated with the $\vec q=0$ order, while for $N$ of multiples of nine ($N=18,27$) the nine sublattices associated with the $\sqrt{3}\times \sqrt{3}$ order in addition. In Fig.1(a), we show the $N$-dependence of the ground-state $m_s^2$-values for several values of the randomness $\Delta$.  In the Heisenberg model sustaining a nonzero $m_s^2$, the spin-wave form, $m_s^2 = m_{s,\infty}^2 + \frac{c_1}{\sqrt{N}}$, is expected for its $N$-dependence. As can be seen from the figure, the data  extrapolated to $N=\infty$ on this form yields negative values for all $\Delta$, indicating that the AF LRO is absent either in the regular or in the random case.

 A nontrivial issue is whether the randomness ever induces the spin-glass (SG) order or not. To examine this possibility, we compute the SG-type spin freezing parameter $\bar q$, defined by $\bar q ^2= \frac{1}{N^2} \sum_{i,j} [\langle  \vec S_i\cdot \vec S_j \rangle ^2]$, $\langle \cdots \rangle $ and $[\cdots ]$ being the ground-state expectation and the sample average,  respectively, and the sum over $i$ and $j$ is taken over all sites of the lattice. The quantity takes a nonzero value when the spin is frozen either in a spatially periodic or random manner \cite{Viet}. As can be seen from Fig.1(b), the randomness tends to enhance the $\bar q$-value somewhat. Yet, linear extrapolation of the data yields $\bar q=0$ within the error bar for any $\Delta$. For $\Delta=1$, although the extrapolation using all available $N$ yields a very small nonzero value $\bar q=0.006\pm 0.002$, the extrapolation using only four largest $N$ yields $\bar q=-0.003\pm 0.006$, indicating the vanishing $\bar q$. Thus, we tend to conclude that the spin freezing does not occur for any $\Delta$, even including $\Delta=1$.

 To get further information, we compute the quantities related to the scalar chirality $\chi_c =\vec S_1\cdot \vec S_2\times \vec S_3$ ($\vec S_1$, $\vec S_2$ and $\vec S_3$ are three corner spins on an upward triangle $t$ on the lattice). While the chirality locally takes a nonzero value because of quantum fluctuations, the chiral freezing parameter $\bar q_\chi$, defined by $\bar q_\chi^2=(\frac{3}{N})^2 \sum_{t,t'} [\langle  \chi_t \chi_{t'} \rangle ^2]$, vanishes in the $N\rightarrow \infty$ limit at any $\Delta$, indicating that the long-range chiral order does not occur, even the random one. Supplemental material gives the details. This observation further supports our conclusion above that the SG-type order does not occur for any $\Delta$. The absence of the chiral order in the non-random case was also reported in ref.\cite{Depenbrock}.

 Next, we study the the finite-temperature properties of the model. In Fig.2(a), we show the temperature dependence of the specific heat per spin $C$ (in units of Boltzmann constant) for several values of $\Delta$. The specific heat of the random model exhibits a $T$-linear behavior $C \simeq \gamma T$ at lower temperatures.
 For $\Delta=0$ and 0.3, the finite-$N$ data exhibit a low-$T$ peak around $T\simeq 0.1$ in addition to the one at a higher temperature. Although the recent calculation for the larger lattice of $N=30$ suggests that this low-$T$ peak might go away in the $N\rightarrow \infty$ limit \cite{Sugiura}, a weaker structure like a hump or a cusp is likely to remain even in the the $N\rightarrow \infty$ limit. By contrast, our data for the random model indicates that even such a weaker low-$T$ anomaly does not appear for $\Delta\geq 0.5$. Hence, some changeover of the low-$T$ property seems to occur when the randomness exceeds a critical value $\Delta_c\simeq 0.4$. 

%
\begin{figure}
 \includegraphics[width=8cm]{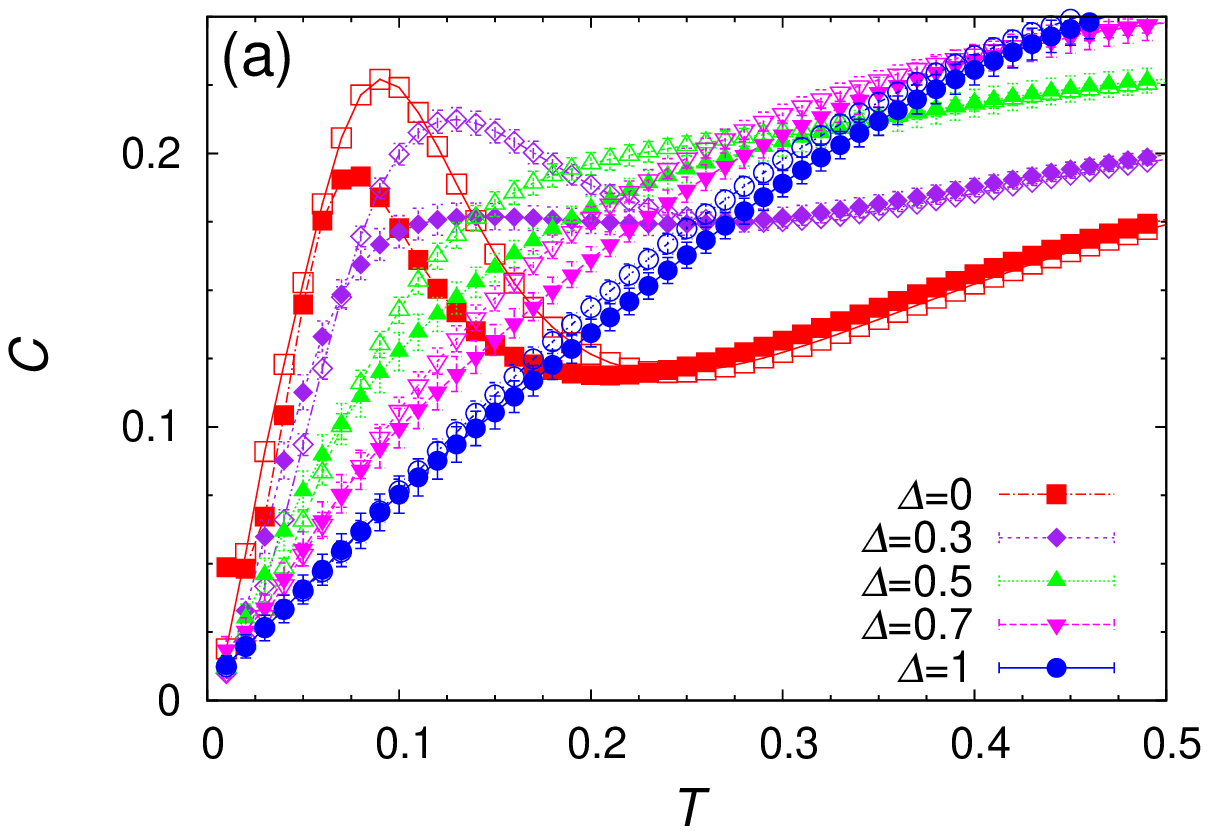}
 \includegraphics[width=8cm]{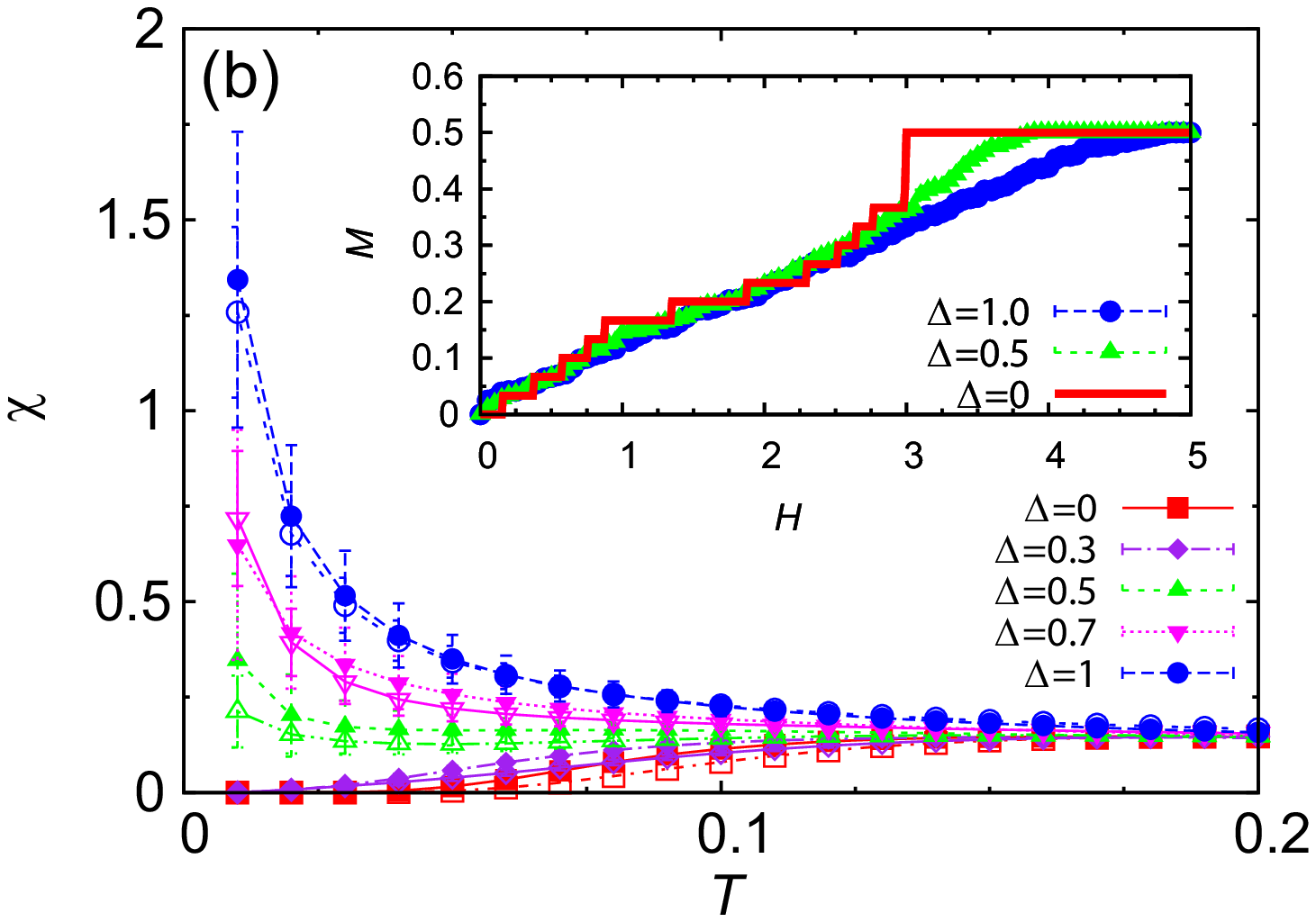}
 \caption{(color online). 
The temperature dependence of (a) the specific heat per spin $C$, and of (b) the uniform susceptibility per spin $\chi$, for several randomness $\Delta$ for $N=12$ (open symbols) and 18 (solid symbols). The inset of (b) represents the $T=0$ magnetization curve for $N=30$.
}
\end{figure}
\begin{figure}
 \includegraphics[width=8cm]{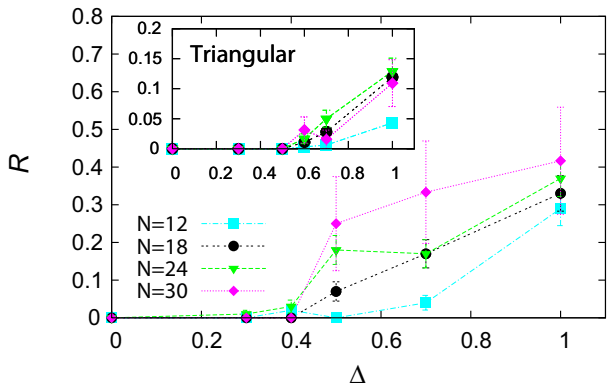}
 \caption{(color online). 
The randomness $\Delta$ dependence of the ratio of samples with triplet grounds states for various even $N$. The inset is the corresponding plot for the $S=1/2$ triangular-lattice model.
}
\end{figure}

 In Fig.2(b), we show the temperature dependence of the magnetic susceptibility per spin $\chi$ for several values of $\Delta$.  While the susceptibility goes to zero in the $T\rightarrow 0$ limit for weaker randomness $\Delta\lesssim 0.3$, it tends to a finite value at $\Delta=0.5$, and diverges obeying the Curie law for stronger randomness $\Delta\geq 0.7$. The appearance of a Curie-like component suggests that a small fraction of free spins ($\sim$ 4\%) are generated at low temperatures.

 In the inset of Fig.2(b), we show the ground-state magnetization per spin $M$ versus an applied field for our largest size $N=30$. It has been known that the non-random model exhibits a plateau-like structure at the $1/3$ of the saturation value (refer also to ref.\cite{Nakano14}), with further additional structures at $1/9$, 5/9 and 7/9 \cite{Nishimoto}. For $\Delta>\Delta_c$, such plateau-like anomalies tend to go away, yielding a near linear behavior similar to the one observed in the corresponding triangular model.

\begin{figure}
 \includegraphics[width=8cm]{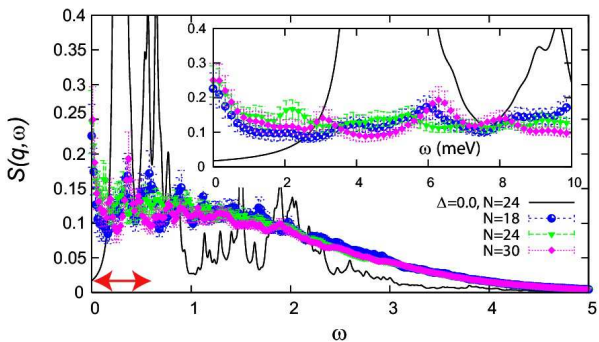}
 \caption{(color online). 
 The dynamical spin structure factor $S(\vec q, \omega)$ at $T=0$ at the $\Gamma$ point plotted versus $\omega$ for the randomness $\Delta =1$, in comparison with that of the non-random model of $\Delta=0$. The broadening factor $\eta$ introduced as in refs.\cite{Gagliano,Lauchli} is set to $\eta=0.02$. The inset is a magnified view of the low-$\omega$ region (indicated by the arrow), the $\omega$-value transformed to that of herbertsmithite with $J=17$ meV.
}
\end{figure}

 The ground state for weaker randomness turns out to be a spin singlet corresponding to the total spin $S=0$ for all even-$N$ samples (for odd $N$, a doublet corresponding to $S=1/2$). As the randomness gets stronger, some fraction of samples has a spin-triplet ground state. In Fig.3, we show the ratio of even-$N$ samples with spin-triplet ground states, $R$, versus $\Delta$. As the randomness gets stronger beyond $\Delta_c\simeq 0.4$, $R$ becomes nonzero, and increases with $\Delta$. For comparison, we show $R$ of the $S=1/2$ triangular-lattice model in the inset, for which $\Delta_c\simeq 0.6$ separates the AF and the random-singlet states. While the triplet tendency is enhanced in the kagome case than in the triangular case,  the two data look similar. Hence, some sort of rapid change, possibly a phase transition, is likely to occur at the critical randomness $\Delta_c\simeq 0.4$ in the ground state of the kagome model. This is corroborated by our data of the specific heat and the susceptibility or the magnetization of Fig.2 where the change of behavior is also observed around $\Delta=\Delta_c\simeq 0.4$.

 The state at $\Delta>\Delta_c$ has properties similar to those of the random-singlet state identified in the corresponding triangular-lattice model, including the $T$-linear specific heat, the gapless susceptibility with a Curie-like component for the stronger randomness, and the near linear magnetization curve without plateau-like structures \cite{Watanabe}. The gapless feature is also consistent with the one suggested for the site-dilution model of ref.\cite{Singh10}. So, we identify the rapid change observed at $\Delta=\Delta_c\simeq 0.4$ as a transition from the randomness-irrelevant QSL state to the randomness-relevant QSL state, the random-singlet (or VBG) state. Our data are not capable of discriminating the former being either the gapped $Z_2$ liquid or the gapless $U(1)$ liquid, but the observation of the vanishing chiral order seems to rule out the chiral QSL state. If the QSL state of the non-random system is the gapped $Z_2$ spin liquid, it would be stable against a weak randomness up to a certain point, and $\Delta_c$ would be a true critical point separating the gapped/gapless behaviors.

 To further characterize each phase, we compute the dynamical spin structure factor $S(\vec q, \omega)$, a quantity accessible by inelastic neutron scattering, following refs.\cite{Gagliano,Lauchli}. The computational details are given in Supplemental material. Its $\omega$-dependence computed for $\Delta=1$ at the $\Gamma$ point $\vec q=(0,\frac{2\pi}{\sqrt{3}d})$ ($d$ the lattice spacing) is shown in Fig.4, together with the one for the non-random case $\Delta=0$ for comparison \cite{Lauchli}. A similar calculation is also made for the $M$ point $\vec q=(0,\frac{\pi}{\sqrt{3}d})$. The intensity at the $\Gamma$-point is somewhat greater than that at the $M$-point, but its $\omega$-dependence is rather similar.

 The size-dependence of the random system turns out to be rather mild. The gapless behavior is observed both at the $\Gamma$ and the $M$ points, without any sharp structure. At $\Delta=1$, the observed intensity exhibits a mild peak at $\omega=0$, with a very broad, almost flat plateau extending to higher $\omega$. The $\omega=0$ peak is not pronounced, the peak height being about twice the intensity at the plateau part. In the non-random case of $\Delta=0$, by contrast, the low-energy peak is much more enhanced, the peak intensity being an order of magnitude greater \cite{Lauchli}. The observed features of the data, a broad $\omega\simeq 0$ peak and almost flat intensity extending to higher $\omega$, especially no gap appearing at any $\vec q$, are qualitatively consistent with the recent inelastic neutron-scattering data on single crystals \cite{Han}. Since the experimental data now available are limited to the low-energy region of $\omega\lesssim 10$meV ($J\simeq 17$meV), we show in the inset a magnified view of the experimentally accessible low-$\omega$ range. The data of the random system appears to resemble the experimental one much more than the one of the non-random system. This observation lends support to the view that the randomness is certainly playing a significant role in herbertsmithite.

 The remaining question is whether herbertsmithite lies at $\Delta<\Delta_c$ or at $\Delta>\Delta_c$. If the former is the case, the QSL state might be a mild modification of that of the non-random system, though the translational symmetry is lost at the Hamiltonian level as soon as $\Delta$ becomes nonzero. If the latter is the case, on the other hand, the QSL state of herbertsmithite would be the random-singlet state. Although the direct contribution from the interlayer Cu$^{2+}$ and the effect of the Dzaloshinskii-Moriya interaction needs to be considered for further comparison, the gapless features observed experimentally in various quantities, the single-crystal $S(\vec q, \omega)$ data in particular, seems to favor the latter scenario, {\it i.e.\/}, the QSL state of herbertsmithite be the random-singlet state. 
Further experiments at higher fields as well as inelastic neutron measurements at higher-$\omega$ are desirable. 

 In summary, we investigated both the zero- and finite-temperature properties of the spin-1/2 bond-random Heisenberg AF on the kagome lattice, to find that the model exhibited beyond a critical randomness a gapless QSL behavior, which was attributed to the random-singlet (or VBG) state. The result might consistently explain the recent experimental data on herbertsmithite.

 The authors are thankful to ISSP, the University of Tokyo for providing us with CPU time. This study is supported by Grants-in-Aid for Scientific Research No. 25247064. The calculation was performed by use of TITPACK Ver.2.


\bigskip\bigskip
\bigskip\bigskip
\noindent
{\bf Supplemental Material}

\subsection*{Chirality-related quantities}
\noindent

We compute both the local scalar-chirality amplitude $\bar \chi$ and the chiral freezing parameter $\bar q_\chi$. The former is defined by
\begin{equation}
\bar \chi = \sqrt{ \frac{3}{N} \left[ \sum_{t} \left\langle \chi_t^2 \right\rangle \right] } , 
\nonumber
\end{equation}
where the sum over $t$ is taken over all $N/3$ upward triangles on the kagome lattice. The chiral freezing parameter is defined by
\begin{equation}
\bar q_\chi = \frac{3}{N}\sqrt{ \left[ \sum_{t,t'} \left\langle  \chi_t \chi_{t'} \right\rangle ^2 \right] } .
\nonumber
\end{equation}
where the sum over $t$ and $t'$ is taken over all $N/3$ upward triangles.
\begin{figure}[ht]
\includegraphics[width=8cm]{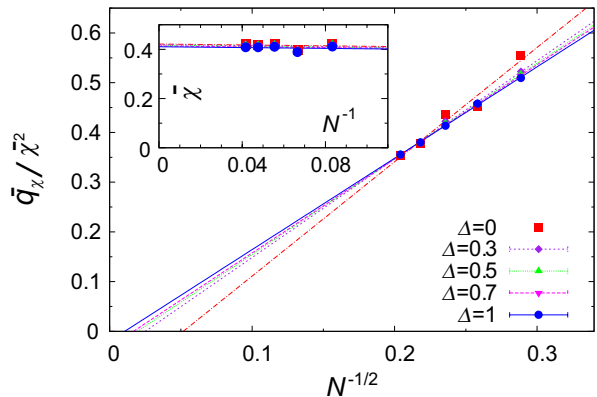}
 \caption{(color online). 
The $T=0$ chiral freezing parameter normalized by its local amplitudes, $\bar q_\chi/\bar \chi^2$, is plotted versus $1/\sqrt{N}$ for several randomness $\Delta$. The lines are linear fits of the data. In  the inset, the local scalar-chirality amplitude $\bar \chi$ at $T=0$ is plotted versus $1/N$.
}
\end{figure}

These quantities are computed by the ED method at $T=0$. The largest system size is limited to $N=24$ for the chirality-related quantities due to the computational load. The computed $\bar \chi$ and $\bar q_\chi$ normalized by its local amplitudes, $\bar q_\chi/\bar \chi^2$, are shown in the figure. As can be seen from its inset, the local scalar-chirality amplitude  becomes nonzero for any $\Delta$, even including $\Delta=0$, indicating that the spin tends to take noncoplanar configuration in the ground state at the local level because of quantum fluctuations. While the scalar chirality takes a nonzero value locally, our data indicates that the long-range scalar chirality, both the uniform and the glass-type random ones, tends to zero for $N\rightarrow \infty$ irrespective of the $\Delta$-value.

\subsection*{Dynamical spin structure factor}
\noindent

The dynamical spin structure factor $S({\bm q}, \omega)$ at $T=0$ is defined by
\begin{equation}
S({\bm q}, \omega) = - \frac{1}{\pi} {\rm Im} \left[ \langle {\bm S}^z(-{\bm q}) \frac{1}{\omega - ({\cal H}-E_0) + i\eta} {\bm S}^z({\bm q}) \rangle \right] ,
\nonumber
\end{equation}
where ${\bm S}^z({\bm q})$ is the Fourier transform of the $z$-component of the spin, and $E_0$ is the ground-state energy. We then compute $S({\bm q}, \omega)$ by the ED method combined with a continued fraction technique [see refs.\cite{Gagliano,Lauchli}]. The broadening factor $\eta$ is set to $\eta=0.02$. 

\end{document}